\documentclass[twocolumn,floats,prb,aps,showpacs]{revtex4}
\usepackage[final]{graphicx}
\usepackage[dvipsnames]{color}
\usepackage{dcolumn}
\usepackage{hhline}
\DeclareGraphicsExtensions{.pdf,.eps}

\begin{document}

\title{ Short to long-range charge-transfer excitations in the zincbacteriochlorin-bacteriochlorin complex: 
a Bethe-Salpeter study}

\author{I. Duchemin$^1$, T. Deutsch$^1$, X. Blase$^2$}

\affiliation{ 
$^1$INAC, SP2M/L$\_$sim, CEA cedex 09, 38054 Grenoble, France, \\
$^2$Institut N\'{e}el, CNRS and Universit\'{e} Joseph Fourier,
B.P. 166, 38042 Grenoble Cedex 09, France.}

\date{\today}

\begin{abstract}
We study using the Bethe-Salpeter formalism the excitation energies of the 
zincbacteriochlorin-bacteriochlorin dyad, a paradigmatic photosynthetic 
complex. In great contrast with standard time-dependent density functional 
theory calculations with (semi)local kernels, charge transfer excitations 
are correctly located above the intramolecular Q-bands transitions found 
to be in excellent agreement with experiment. Further, the asymptotic 
Coulomb behavior towards the true quasiparticle gap for charge transfer 
excitations at long distance is correctly reproduced, showing that the 
present scheme allows to study with the same accuracy intramolecular and 
charge transfer excitations at various spatial range and screening environment 
without any adjustable parameter.
\end{abstract}

\pacs{71.15.Qe,78.67.-n,78.40.Me}
\maketitle


Photoinduced charge transfer excitations, namely the jump upon photon absorption of an
electron from a donor to an  acceptor site, is a fundamental process that governs 
photosynthetic processes in plants and bacterias, \cite{photosynthesis} or the quantum
efficiency in organic or hybrid photovoltaic cells. \cite{photovoltaics} Such non-local
excitations are also an important current theoretical issue since it is now well recognized
that the time-dependent density functional theory (TDDFT) \cite{tddft} encounters severe 
problems to describe such excitations when  standard (semi)local kernels, or even hybrid 
kernels mixing some amount of exact exchange, are being used. \cite{Dreuw04} Besides organic
systems, similar problems have been identified in the case of extended Wannier excitons 
in semiconductors where the large effective excitonic radius leads to a weak 
average overlap between the hole and the electron.  \cite{Botti04} 

\begin{figure}
\begin{center}
\includegraphics*[width=0.4\textwidth]{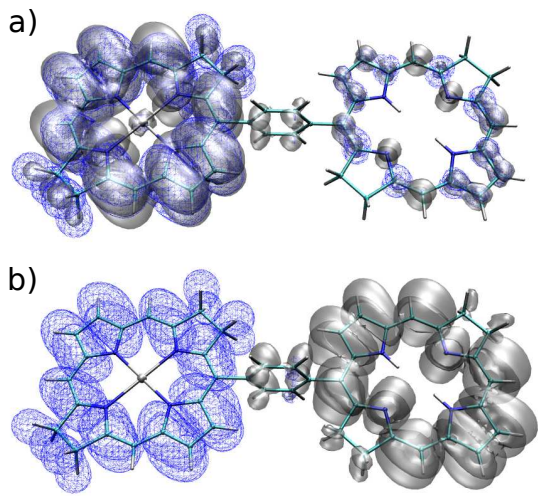}
\caption{ (Color online) Symbolic representation of the (1,4)-phenylene-linked 
zincbacteriochlorin-bacteriochlorin complex. Isocontour representation of the (grey) 
hole-averaged electron and (blue) electron-average hole distributions (see text) for
the lowest (a) ZnBC $Q_x$ excitation and (b) ZnBC $\rightarrow$ BC charge-transfer
excitation. }
\label{fig1}
\end{center}
\end{figure}

The bacteriochlorin molecule is closely related to the magnesium-containing bacteriochlorophyll
system.  Due to its importance as a paradigmatic photosynthetic complex, and as one of the earliest 
charge-transfer system for which the TDDFT difficulties have been unraveled and discussed, 
\cite{Dreuw04} the zinc-bacteriochlorin/bacteriochlorin (ZnBC-BC) complex 
(see Fig.~\ref{fig1}) has been studied by a variety of approaches, including TDDFT
with local, \cite{Dreuw04,Yamaguchi02,Kobayashi06} hybrid, \cite{Kobayashi06} Coulomb attenuated 
hybrid \cite{Kobayashi06} functionals, constrained $\Delta$SCF DFT calculations, \cite{Wu05} 
and quantum chemistry many-body wavefunctions techniques such as a combination of $\Delta$SCF DFT
and single excitation configuration interaction (CIS) technique, \cite{Dreuw04} or a more elaborate 
CIS(D) approach including various scaled perturbative double-excitation correlation corrections. \cite{Rhee07}

In two recent studies, \cite{Lastra11,Blase11b} charge transfer (CT) excitations in small
donor/acceptor complexes, combining tetracyanoethylene  with acene derivatives, were 
studied with the $GW$ approximation and Bethe-Salpeter (BSE) equation within 
many-body perturbation theory. \cite{Rocca10} 
Excellent agreement with gas phase experiments \cite{Hanazaki72} was obtained 
for the lowest CT excitation energy with a mean absolute error of about 0.1 eV.
\cite{Blase11b}  Such an accuracy compared well with recent TDDFT calculations with optimized 
range-separated functionals, \cite{Stein09} while TDDFT calculations with standard PBE or
even non-local B3LYP kernels \cite{PBE,B3LYP} were shown to lead to discrepancies of several 
eV with CT states located at much too low energy. \cite{Stein09} 

In this work, we study the optical absorption spectrum of the ZnBC-BC complex using
the $GW$-BSE many-body perturbation theories. We show that intramolecular excitations are
in excellent agreement with experiment, and that the charge-transfer excitations are correctly 
located above the intramolecular Q-bands transitions, in great contrast 
with TDLDA calculations which locate the charge-transfer excitations about 1.7 eV below the 
$GW$-BSE value. It is further shown that the $GW$-BSE framework correctly reproduces the 
long-range energy behavior of the charge transfer excitations, a feature shared only with 
the scaled  CIS(D) correlated quantum chemistry approach. The importance of going beyond 
the Tamm-Dancov approximation is further discussed. 

Our calculations are based on a recently developed auxiliary even-tempered Gaussian-basis 
\cite{Cherkes09} implementation of the $GW$ \cite{Hedin65,Hybertsen86,Godby88,Onida02}
and BSE \cite{Sham66,Strinati82,Rohlfing98,Benedict98,Albrecht98} formalisms, the 
{\sc{Fiesta}} package, implementing contour deformation techniques for the 
correlation contribution to the self-energy. \cite{Blase11a,Blase11b} The 
needed two-body operators such as the susceptibilities, bare/screened Coulomb potentials 
and self-energies, are expressed on an atom-centered Gaussian-basis containing six 
\textit{exp}$(-\alpha {\bf r}^2)$ gaussian for the radial part of each (\textit{s}, \textit{p}, 
\textit{d})-channels, with an even tempered distribution of localization coefficient 
$\alpha$ ranging from 0.1 to 3.2 a.u. Such a basis derives from previous studies 
\cite{Blase11a,Ma10,Faber11a} but with additional diffuse orbitals.

The needed starting single-particle states are taken to be the Kohn-Sham DFT/LDA eigenstates 
generated by the {\sc{Siesta}} package \cite{siesta} using a large triple-zeta plus double 
polarization basis (TZDP). \cite{pseudo}  It was shown recently that the combination 
of a TZDP basis for describing Kohn-Sham occupied and virtual states, with the above described
auxiliary bases, was leading to charge-transfer excitation energies in excellent agreement with 
planewave $GW$-BSE calculations. \cite{Blase11b} For the isolated ZnBC and BC molecules, 
for which experimental data are available, \cite{exp1,exp2} we relax the molecules at the 
all-electron B3LYP 6-311G(d) level. \cite{gaussian} However, for the ZnBC-BC bridged complex,
we adopt the geometry  provided in Ref.~\onlinecite{Dreuw04}, and used in subsequent studies, 
\cite{Kobayashi06,Wu05,Rhee07} as a mean to properly compare to other computational approaches.

In the present many-body framework, one first obtains accurate quasiparticle energies (occupied 
and virtual electronic energy levels) by using the $GW$ formalism. In the case of gas phase
organic molecules such as fullerenes, pentacene, PTCDA or porphyrins, it was shown in several 
recent studies \cite{Blase11a,Tiago08,Palummo09,Sahar12} that the standard single-shot 
$G_0W_0$ calculations based on input Kohn-Sham eigenstates with standard local functionals tends 
to yield too small HOMO-LUMO gaps, even though already much better than the starting Kohn-Sham 
values. Better results could be consistently obtained for the quasiparticle spectrum of small 
molecules either by starting from Hartree-Fock 
\cite{Blase11a,Hahn05,Rostgaard10,Ke11}, or hybrid functionals \cite{Marom11} eigenstates, or 
by performing a full self-consistency on the quasiparticle eigenstates,  
\cite{Rostgaard10,Ke11} or a limited self-consistency on the eigenvalues only, 
\cite{Blase11a,Blase11b,Faber11a,Faber11b,Sahar12} allowing to remove the dependency of the 
final quasiparticle energies on the starting point eigenstates. This is such an approach that 
we adopt in the present study, showing here below as in Ref.~\onlinecite{Blase11b} that excellent 
agreement with experiment can be obtained for the excitation energies.

In an optical absorption spectrum, the quasiparticle gap is reduced by the electron-hole 
interaction.  In the present MBPT formalism, this excitonic interaction is accounted for 
by the BSE equations.  Namely, the neutral excitation energies can be obtained as the 
eigenvalues of the BSE $H^{e-h}$ Hamiltonian 
\cite{Sham66,Strinati82,Rohlfing98,Benedict98,Albrecht98} expressed in the 
$\phi_{i}^e(r) \phi_j^h(r')$ product basis of the unoccupied $\phi_{i}^e$ 
and occupied  $\phi_j^h$  single-particle states. Of importance for the upcoming discussion, 
we write the so-called direct term involving the (statically) screened 
Coulomb potential $W({\textbf r},{\textbf r'})$, noticing that hole and electron states are 
not taken at the same space position:

$$
  H^{direct}_{ij,kl} = -  \int d{\textbf r} d{\textbf r'} \phi_i^e({\textbf r}) \phi_j^h({\textbf r'}) 
        W({\textbf r},{\textbf r'}) \phi_k^e({\textbf r}) \phi_l^h({\textbf r'}),
$$

\begin{center}
\begin{table*}
\begin{tabular}{l|cccc|cc|c}
\hline
                            &   \multicolumn{4}{c|}{TD-DFT (TDA)}    & \multicolumn{2}{c|}{$GW$-BSE}  & Exp.$^{c,d}$  \\ 
\hline
Transitions                 & BLYP$^{a/b}$ &   LDA &  B3LYP $^b$ & CAM-B3LYP$^b$ &  TDA & full &  \\
\hline
                            &       \multicolumn{7}{c}{Isolated BC and ZnBC monomers}                                             \\
$\pi$-${\pi}^*$ $Q_x$ ZnBC  &  2.07/2.04  & 2.09  &  2.05  & 1.87 & 1.94 & 1.59  & 1.65  \\
$\pi$-${\pi}^*$ $Q_x$ BC    &  2.10/2.10  & 2.10  &  2.12  & 1.92 & 1.90 & 1.63  & 1.6  \\
$\pi$-${\pi}^*$ $Q_y$ BC    &  2.39/2.40  & 2.45  &  2.54  & 2.53 & 2.38 & 2.24  & 2.3  \\
$\pi$-${\pi}^*$ $Q_y$ ZnBC  &  2.44/2.43  & 2.47  &  2.60 & 2.59 & 2.29 & 2.27  & 2.2  \\
                            &       \multicolumn{7}{c}{ZnBC-BC phenylene-bridged complex}                                         \\
$\pi$-${\pi}^*$ $Q_x$ ZnBC & 2.05/2.05  & 2.08  &  2.41  & 1.89 &  1.99  & 1.69  &      \\
$\pi$-${\pi}^*$ $Q_x$ BC    & 2.09/2.09  & 2.10  &  2.43 & 1.93 & 2.04  & 1.73   &      \\
$\pi$-${\pi}^*$ $Q_y$ BC    & 2.38/2.37  & 2.46  &  2.79 & 2.49 & 2.47  & 2.35   &      \\
$\pi$-${\pi}^*$ $Q_y$ ZnBC & 2.42/2.42  & 2.44  &   2.83 & 2.54 & 2.49  & 2.34   &      \\
CT ZnBC $\rightarrow$ BC    & 1.33/1.33  & 1.25  &  1.96  & 2.87 & 2.95  & 2.95   &      \\
CT BC $\rightarrow$ ZnBC    & 1.46/1.46  & 1.39  &  2.12  & 3.04 &  3.13  & 3.13   &      \\
\hline
\end{tabular}
\caption{Calculated singlet transition energies (in eV) for the isolated monomers and the (1,4)-phenylene-linked 
ZnBC-BC complex.  The present TD-LDA (TDA),  $GW$-BSE (TDA) and $GW$-BSE (full) results are compared to previous 
TD-DFT  calculations with the BLYP and B3LYP functionals. (TDA) means Tamm-Dancov approximation, while 
(full) means full diagonalization mixing resonant and antiresonant transitions.  
$^a$Ref.~\onlinecite{Dreuw04}.  $^b$Ref.~\onlinecite{Kobayashi06}. $^{c,d}$Refs.~\onlinecite{exp1,exp2}. }
\label{tablebse}
\end{table*}
\end{center}


\noindent
As compiled in Table~\ref{tablebse}, we first verify that our $GW$-BSE low-lying $Q_x$ and $Q_y$
transition \cite{gouterman} energies for the isolated monomers fall within 0.1 eV of the available
experimental values.  Consistently with previous observations, \cite{Gruning09,Ma10} we find that 
the diagonalization of the full BSE Hamiltonian leads to a red shift which can be as large as 0.35 eV 
as compared to the $GW$-BSE results in the Tamm-Damcov (TDA) approximation, bringing the calculated 
transitions in excellent agreement with experiment. Such a good agreement with available experimental
data can be taken as an indication of the reliability of the present formalism and of the specific 
implementation aspects. 

Concerning the TDLDA calculations, performed with the same Kohn-Sham states and auxiliary basis, 
we observe that the agreement with experiment is also satisfactory, even though not as good as the 
BSE values. Our calculated TDLDA results compare very well with former TD-BLYP calculations. 
\cite{Dreuw04,Kobayashi06} Again, this allows to verify that the pseudopotential approximation and
the bases we use do not lead to significant errors as compared to all-electron calculations with 
standard quantum chemistry basis. For intramolecular transitions with large overlap between hole and
electron states, the TDDFT approach with standard kernels is a reliable framework.

We now come to the central point of this study, namely the ZnBC-BC CT excitation energies.
The charge-transfer nature of a given transition can be easily identified by analyzing the weight 
of the two-body BSE eigenstates on the $\phi_{lumo}^e$ and  $\phi_{homo}^h$ one-body eigenstates 
localized on either molecule.  The expectation value of the electron/hole density 
operator $\delta({\bf r}-{\bf r}_e/{\bf r}_h)$ on a given two-body BSE eigenstate allows further 
to build the corresponding hole/electron-averaged electron/hole distribution. Such a representation 
is provided in Fig.~\ref{fig1}(b) for the ZnBC$\rightarrow$BC charge-transfer state obtained at 
the $GW$-BSE (full) level. The charge-transfer nature of such an excited state is apparent.  

As shown in Table~\ref{tablebse}, TD-LDA severely underestimates the CT excitation energies 
which are found to lie significantly below the intramolecular Q-bands. Identical conclusions 
were found with the BLYP functional. \cite{Dreuw04,Kobayashi06} Such a behavior can 
be understood from the analysis of the TDDFT electron-hole coupling terms which vanish for
transitions between non-overlapping electron and hole states in the case of (semi)local 
exchange-correlation kernel. One is then left with the non-interacting diagonal part, namely 
the too small Kohn-Sham HOMO-LUMO gap. Such a situation is slightly improved with the B3LYP
functional due to its 20$\%$ of exact exchange, but with a residual error as large as 1 eV. 
This cancellation of the interacting term for non-local charge-transfer excitations does not occur 
in the BSE formalism since in the $H^{direct}_{ij,kl}$ interaction term, hole and electron states 
are not taken at the same position in space and are connected by the non-local screened Coulomb
potential $W({\bf r},{\bf r'})$. 

The $GW$-BSE  CT excitation energies lie  well above the intramolecular Q-bands, 
consistently with TDDFT calculations with optimized (parametrized) CAM-B3LYP functionals 
\cite{Kobayashi06} and CIS(D) calculations \cite{Rhee07} (see analysis below). 
Concerning the CAM-B3LYP results, with CT states located less than 0.1 eV below 
the $GW$-BSE one, to be compared to the $\sim$ 1.7 eV discrepancy with TDLDA, 
the agreement is very good considering that the needed
($\alpha,\beta$) scaling parameters \cite{Yanai04} have been trained on a very different set of 
molecules, the so-called G2 set, and on very different properties, namely ionization energies. 
These results confirm the agreement already found between $GW$-BSE and a TDDFT calculation with 
another optimized range-separated functional  (the BNL functional \cite{Livshits07}) 
in the study of small TCNE-acenes complexes, both calculations coming in close agreement with 
experiment for CT states. \cite{Stein09,Blase11b} 

\begin{figure}
\begin{center}
\includegraphics*[width=0.48\textwidth]{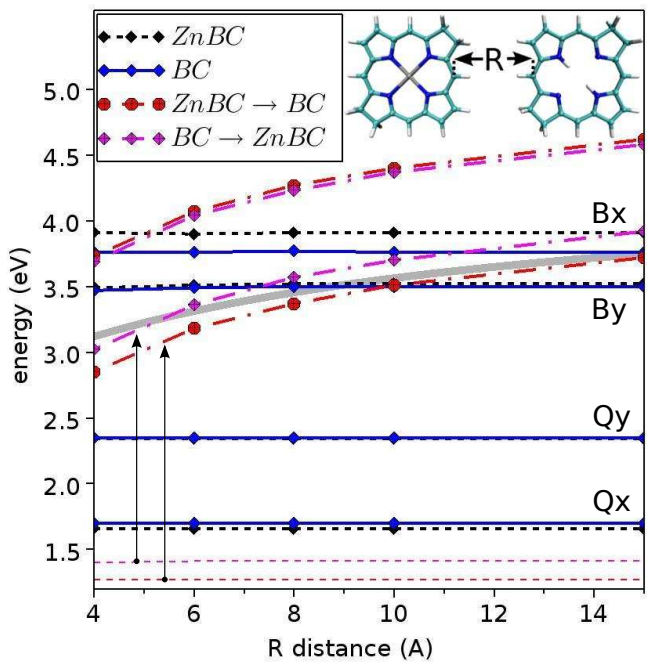}
\caption{ (Color online) $GW$-BSE excitation energies as a function of the ZnBC to BC distance R 
in the absence of the phenylene bridge as indicated on the atomic representation. The Inset indicates
the intramolecular or charge-transfer nature of the represented $GW$-BSE transitions.  
The lowest ZnBC $\rightarrow$ BC CT excitation energy is compared to the ($E^{GW}_{Gap}$ - $e^2$/D) 
``Mulliken" limit (thick grey line), with $E^{GW}_{Gap}$ the dimer $GW$ quasiparticule band gap 
in the large D limit, where D is the distance between the two molecule centers. For sake of comparison,
the TDLDA values for the two low lying CT excitons are indicated by the two non-dispersive dotted lines 
below 1.5 eV.  The two vertical arrows indicate the difference between the TDLDA and $GW$-BSE values. 
For completness, the higher lying intramolecular Soret B-transitions and CT excitations, not detailed
in the Table, are also reproduced.}
\label{asymptotics}
\end{center}
\end{figure}

The ability of the BSE approach to describe charge-transfer excitations can be further illustrated 
by studying the long range limit where the BC and ZnBC units are separated by removing the 
phenylene bridge.  The distance R between the two monomers in the unbridged model 
dimer is defined in Fig.~2a (Inset).  R $\sim$ 5.84~\AA~is the corresponding distance between the two 
monomers in the phenylene-bridged complex.  Our results are represented in Fig.~\ref{asymptotics}. 
As expected, intramolecular charge transfer excitations are independent of the distance between the 
molecules.  On the contrary, in the large distance limit, the CT exciton binding energies
are found to scale like 1/D, where D is the distance between the $R_1$ and $R_2$ molecule centres, 
as indicated in Fig.~\ref{asymptotics}b (thick grey line). 
In the large D limit, the direct interaction terms converge to:
$\;\; H^{direct}_{ij,kl}  \simeq   -  W({\textbf R}_1,{\textbf R}_2) Q^e_{ik} Q^h_{jl}$, 
where $W({\textbf R}_1,{\textbf R}_2)$ reduces to the bare $1/|{\textbf R}_1-{\textbf R}_2|$
Coulomb potential for the two molecules in the vaccuum \cite{TDSEX} and $Q^h_{jl}=<\phi_i^e|\phi_k^e>$.
With $Q^e_{ik}= \delta_{ik}$ and $Q^h_{jl} = \delta_{jl}$, there is no mixing with higher subbands, 
so that the lowest CT state is primarily composed of the HOMO and LUMO eigenstates in the long distance
limit, as verified by analyzing the BSE eigenstates. 

Such (1/D) behavior of the CT excitonic
binding energy is the correct asymptotic limit of an electrostatic interaction between opposite charges.
In this limit, the CT excitation energies correctly converge towards the non-interacting 
(diagonal) part of the BSE Hamiltonian, namely the true quasiparticle gap between the concerned
hole and electron states as given by the $GW$ calculation. 
\textit{ The ability of the $GW$-BSE approach to correctly describe without adjustable parameter both 
intramolecular and  short to long-range charge transfer excitations energies, both in finite (molecular)
and extended (infinite solids) systems}, \cite{bulk} \textit{ is a important feature as discussed here
below by comparison with previous studies. }

As emphasized above, TD-DFT calculations with local kernels cannot reproduce the long-range 
electron-hole interaction: the CT energies remain constant and equal to the Kohn-Sham band gap.
This is now well documented \cite{Dreuw04,Yamaguchi02,Kobayashi06} and we clearly verify that 
point with our TDLDA data (see dotted lines below 1.5 eV in Fig.~\ref{asymptotics}).  As a dramatic 
improvement over local kernels, the TD-CAM-B3LYP approach \cite{Kobayashi06} provides a much 
better agreement with $GW$-BSE calculations in the short-range limit of the bridged dimer 
(see above).  However, as already noticed in Ref.~\onlinecite{Kobayashi06}, the ($\alpha+\beta$)/D 
long-range behavior,  with ($\alpha+\beta$)=0.65, \cite{Yanai04} cannot reproduce the correct 1/D  
long-range limit. Similarly, the asymptotic limit of the TD-B3LYP approach would reduce to 0.2/D,
thanks to the 20$\%$ of exact exchange. This discussion clearly underlines the difficulty of working
with fixed parameters functionals performing equally well at any range
and in any screening environment (vaccuum, solvant, etc.)  \cite{TDSEX}

As a final comparison, we now analyse the difference with the hybrid single configuration 
interaction (CIS) approach proposed in the seminal Ref.~\onlinecite{Dreuw04}. In the TDA 
approximation, valid for CT states in the long-range limit, CIS is equivalent to time-dependent 
Hartree-Fock showing the correct 1/D asymptotic limit for charge transfer excitations, but with
an incorrect asymptotic limit given by the donor-HOMO to acceptor-LUMO gap calculated in the absence 
of correlations. The neglect of correlation in TD-HF was corrected by a distance independent 
term obtained from DFT $\Delta$SCF calculations at large distance. For $R$=5.85~\AA, the R 
distance in the true bridged configuration, this scheme was shown to lead a low-lying 
ZnBC $\rightarrow$ BC~CT excitation energies of 3.75~eV. This is $\sim$ 0.6~eV larger than 
our $GW$-BSE 3.16~eV value found for the model dimer at the same distance.

As analyzed in Ref.~\onlinecite{Rhee07}, the assumption of a distance independent correlation 
correction to CIS leads to somehow too large excitation energies. This problem was addressed
in a recent study \cite{Rhee07} where scaled (parametrized) perturbative CIS(D) double excitation
corrections to CIS were introduced to account for correlations, \cite{scaling} reducing the 
discrepancy with our BSE calculations to about 0.25 eV.~\cite{cisdfig} While differences in 
basis size may possibly explain part of the small residual discrepancy, \cite{cisdfig} it 
remains that the comparison of quantum chemistry post-Hartree-Fock correlated techniques, such 
as second-order M{\o}ller-Plesset (MP2) methods (see Note~\onlinecite{scaling}), with the present 
$GW$-BSE approach, is a current challenge, with MP2 techniques facing scaling and divergency 
problems for extended systems with large polarizability, \cite{Gruneis10} while the $GW$-BSE 
approach on the contrary still needs futher validation for finite size molecules. The present 
study clearly aims at contributing to that important goal.

In conclusion, we have performed an \textit{ab initio} many-body $GW$-BSE perturbation theory 
analysis of the optical transition energies for the ZnBC-BC complex. For the isolated monomers, 
the calculated $GW$-BSE Q-bands transition energies are found to be in remarkable agreement 
with experiment, in particular if one goes beyond the Tamm-Dancov approximation. In the case of 
the ZnBC-BC complex, the $GW$-BSE calculations correctly locate the charge-transfer excitations 
above the monomer Q-bands transitions. Our CT excitation energies are found to be within less 
than 0.1 eV from parametrized TDDFT-CAM-B3LYP values for the short range charge transfer in the 
bridged configuration, but only the  $GW$-BSE formalism can reproduce the correct long distance
asymptotic limit.  The possibility to study on the same footing, namely without any system-dependent 
parameter, charge tranfer excitation energies at various ranges, both in the vaccuum or in a 
screening environment (solid, solvant, etc.), opens the way to important developments in the 
study of charge-transfer excitations and energy transfer processes at stake in photosynthetic 
processes or in organic photovoltaic cells.

\textbf{Acknowledgements.}
I.D. acknowledges funding from the CEA ``Eurotalent" program.  The authors are indebted to 
C.~Faber, C.~Attaccalite, V.~Ol\'{e}vano for many suggestions and critical readings of our 
manuscript.



\begin{thebibliography}{150bliography}
\bibitem{TDSEX}
The BSE approach with  static screening is formally equivalent to time-dependent Hartree plus 
screened exchange. As such, the screening automatically  sets the range and strength of the
effective Coulomb interaction. In particular, the correct $1/({\epsilon_M}D)$ long-range 
limit would be correctly described within $GW$-BSE in a medium with macroscopic dielectric 
constant $\epsilon_M$ (e.g. in solvant conditions).
\bibitem{scaling}
This approach is described in Ref.~\onlinecite{Rhee07} as an ``excited state analogues of 
ground state scaled second-order M{\o}ller-Plesset (MP2) methods". The word ``scaled" indicates that the 
implementation relies on scaling parameters educated on various molecules, excluding the ZnBC-BC complex.
\bibitem{cisdfig}
Value estimated from Fig.~9 of Ref.~\onlinecite{Rhee07}.  Calculations in Ref.~\onlinecite{Rhee07} were performed 
with a 6-31G* basis, namely a DZP basis smaller than the TZDP basis used here.
\end{thebibliography}
\end{document}